\documentclass[prr,aps,twocolumn,showpacs,longbibliography,superscriptaddress,10pt]{revtex4-1}

\usepackage{amsmath}
\usepackage{amsfonts}
\usepackage{amssymb}
\usepackage{graphicx}
\usepackage{color}
\usepackage{soul}
\usepackage{xcolor}
\usepackage{footmisc}
\usepackage{braket}
\usepackage{todonotes}

\allowdisplaybreaks

\begin{document}
\title{Platform tailored co-design of gate-based quantum simulation}

\author{Kushal Seetharam}
\affiliation{Department of Electrical Engineering, Massachusetts Institute of Technology, Cambridge, MA, 02139, USA}
\affiliation{Department of Physics, Harvard University, Cambridge, MA, 02138, USA}
\author{Dries Sels}
\affiliation{Department of Physics, New York University, New York, NY, 10003, USA}
\affiliation{Center for Computational Quantum Physics, Flatiron Institute, New York, NY, 10010, USA}
\author{Eugene Demler}
\affiliation{Institute for Theoretical Physics, ETH Z{\"u}rich, 8093 Z{\"u}rich, Switzerland}

\date{\today}

\begin{abstract}
	
The utility of near-term quantum computers and simulators is likely to rely upon software-hardware co-design, with error-aware algorithms and protocols optimized for the platforms they are run on. Here, we show how knowledge of noise in a system can be exploited to improve the design of gate-based quantum simulation algorithms. We demonstrate this co-design in the context of a trapped ion quantum simulation of the dynamics of a Heisenberg spin model. Specifically, we derive a theoretical noise model describing unitary gate errors due to heating of the ions' collective motion, finding that the temporal correlations in the noise induce an optimal gate depth. We then illustrate how tailored feedforward control, best applied at this optimum, can be used to partially mitigate unitary gate errors and improve the simulation outcome. Our results provide a practical guide to the co-design of gate-based quantum simulation algorithms.

\end{abstract}

\maketitle

\section{Introduction}
\label{sec:Intro}

Large-scale fault-tolerant quantum computers have the potential to catalyze progress in physics and material science, chemistry and drug development, as well as optimization and machine learning~\cite{Preskill2018, Georgescu2014}. In the near-term, noisy intermediate-scale quantum (NISQ) technology may still demonstrate a quantum advantage over classical computers for certain tasks. The first useful task of NISQ computers likely to demonstrate a quantum advantage is the simulation of quantum dynamics~\cite{Childs2018, Preskill2018, quantumLeap2019}. Digital quantum simulation, accomplished by discretizing the dynamics into several gates, is a flexible approach with controllable error that can improve our understanding of spin systems~\cite{Lanyon2011, Salathe2015}, quantum chemistry~\cite{Kassal2008,OMalley2016}, biochemistry~\cite{Sels2020}, and high energy physics~\cite{Hauke2013, Martinez2016}. A common challenge in all such gate-based quantum simulation is to optimize the quantum circuit implementing the algorithm for a particular NISQ platform. Specifically, the discretization error in the algorithm is reduced by increasing the number of gates, while hardware noise in the system causing decoherence leads to error that typically worsens as the number of gates increases. To achieve the best performance of the algorithm, we must therefore determine both the optimal number of gates and the optimal parameters for these gates in order to account for noise. The focus of this work is to provide insight into these questions which lie at the heart of software-hardware co-design of gate-based quantum simulation.

More generally, understanding the principles of co-design and error-mitigation is essential to realize the potential of quantum computers, as hardware noise usually wipes out the effects responsible for quantum advantages~\cite{Ladd2010}. Even fault-tolerant quantum computers of the future will rely on the characterization and mitigation of noise. The existence of a fault tolerance threshold is only rigorously defined when errors are assumed to be independent; this Markovian idealization is only true when spatial and temporal correlations in the noise die off quickly~\cite{Nielsen2000, Ng2009, Preskill2012}. The magnitude of independent errors, in turn, affects the resource cost of the system, with noisier systems requiring a larger overhead of physical qubits per logical qubit. For near-term noisy, intermediate-scale platforms, characterizing the noise in a system is even more critical. Practical applications will require the co-design of protocols optimized to different hardwares. Indeed, understanding the nature of noise in a system can enable tailored quantum-control and error mitigation that improves desired performance metrics~\cite{Paz_Silva_2016, Layden2018}. In some cases, noise can even be exploited as a feature of the system to simulate the dynamics of complicated many-body models~\cite{Stannigel2014}.

\begin{figure}[t!]
	\centering
	\includegraphics[clip, width=0.99\columnwidth]{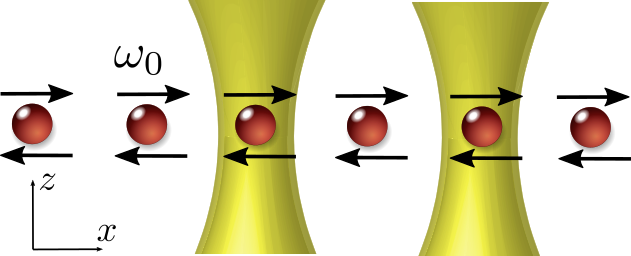}
	\caption{Chain of trapped ions collectively moving at frequency $\omega_{0}$ in the $x$ direction. Ions explore different parts of the beam waist of lasers (yellow) that apply unitary gates, thus accumulating an incorrect phase.}
	\label{fig:ionSchematic}
\end{figure}

The three categories of error in quantum computations and simulations are measurement errors, incoherent errors, and coherent errors. The first, measurement error, arises as quantum observables have inherent uncertainty and hence their expectation value can only be determined with a certainty set by the number of measurement samples. The second, incoherent error, arises from coupling between the qubits and their environment, with these interactions causing the internal state of the qubit to change. The last, coherent error, occurs when a desired unitary transformation of the system imparts an angle different than intended. These unitary errors are the focus of this work and arise due to limitations of the platform's analog control hardware or the dynamics of the physical qubits~\cite{Rines2019}. Over the course of a quantum computation or simulation, such unitary errors accumulate and dephase the system state, killing the coherent effects responsible for a quantum advantage and degrading the fidelity of any simulation. In trapped ions, for example, one dominant source of decoherence is the ions' collective motion, which is thermally excited due to electric field fluctuations from trap electrodes. While the internal qubit states of the ions are not directly affected by this motion, any quantum gate applied via individually addressed lasers imparts an erroneous phase to the qubit states, as depicted in Fig.~\ref{fig:ionSchematic}. We can understand the source of this noise as the phonon mode associated with the center of mass of the chain having an energy that undergoes diffusion due to heating from electric field fluctuations. This slow diffusion of the phonon in energy space causes the unitary error to be non-Markovian, with correlations arising between gates applied at different times during an experiment. 


In this work, we demonstrate how to exploit knowledge of the noise underlying a system to optimize gate-based quantum simulations. To provide an example, we do so in the context of simulating the dynamics of a Heisenberg spin model in a system of trapped ions. We first introduce the quantum simulation task and associated gate-based algorithm. Then, we derive a theoretical noise model describing unitary errors from thermal ion motion in trapped ion systems and provide a protocol to experimentally extract the latent variable underlying the model. We discuss how temporal correlations in the noise induce an optimal gate depth of the quantum simulation circuit. These correlations cause the error in the simulation arising from motional noise to accumulate as the gate depth is increased, while the Trotter error associated with discretization of the time-evolution decreases as the gate depth is increased. The competition of these two errors induces an optimal gate depth.

Next, we provide a platform-independent framework for optimal feedforward control of unitary gate errors, which involves applying gates with angles that are modified to compensate for the predicted noise in the system. We illustrate the utility of feedforward control in the trapped ion implementation of simulating the Heisenberg Hamiltonian, showing that feedforward control partially mitigates both discretization error and decoherence error in the simulation output. 

Our work provides three results that are generally applicable to the co-design of gate-based quantum algorithms beyond the discussed simulation task: \textit{(i)} the understanding that non-Markovian correlations are the root cause of decoherence and the subsequent limitation on gate depth in any platform where unitary errors are the dominant noise, \textit{(ii)} a method to optimally leverage noise characterization to mitigate unitary gate errors via feedforward control, and \textit{(iii)} an accurate model of unitary gate errors arising from thermally-excited ion motion in trapped ion systems.

\section{Hamiltonian simulation}
\label{sec:HamSim}

Simulating the quantum dynamics of a system is a natural application of digital quantum computers and analog quantum simulators, and is likely to be the first problem of practical interest where a quantum advantage over classical computers is demonstrated on near-term quantum platforms~\cite{Childs2018}. The goal is to simulate the time-evolution, $\hat{U}\left(t\right)=\exp\left(-i\hat{H}t/\hbar\right)$, of a system whose dynamics are generated by a Hamiltonian $\hat{H}$. Here, we focus on the Heisenberg Hamiltonian,
\begin{equation}\label{eq:HeisenbergHam}
\hat{H}=\sum_{i,j}J_{ij}\hat{\mathbf{S}}_{i}\cdot\hat{\mathbf{S}}_{j}+\sum_{i}h_{i}\hat{S}_{i}^{x}
\end{equation}
which is a paradigmatic spin model that describes the magnetic properties of many insulating crystals~\cite{Dagotto1994}, appears in the study of thermalization in quantum systems~\cite{Nandkishore2015, Luitz2015, Pal2010}, and describes the essential physics underlying nuclear magnetic resonance (NMR) spectroscopy~\cite{Sels2020}. 

Many near-term quantum algorithms and simulations focus on the task of estimating the expectation value of some observable after time-evolution, with the value of such observables often being less susceptible to noise than the full system state~\cite{Temme2017}. In this vein, we benchmark the quality of gate-based quantum simulation of Eq.~\eqref{eq:HeisenbergHam} with the spectrum simulation task discussed in Ref.~\cite{Sels2020} and experimentally demonstrated in Ref.~\cite{Seetharam2023}. This algorithm uses time-evolution under the Heisenberg Hamiltonian, Eq.~\eqref{eq:HeisenbergHam}, to compute the NMR spectrum
\begin{equation}\label{eq:Spectrum}
A\left(\omega\right)=\text{Re}\int_{0}^{\infty}dt\cdot e^{i\omega t-\gamma t}S\left(t\right)
\end{equation}
where 
\begin{align}
S\left(t\right) & =\braket{\hat{S}_{\text{tot}}^{z}\left(t\right)\hat{S}_{\text{tot}}^{z}}\\
& =\text{Tr}\left[e^{i\hat{H}t}\hat{S}_{\text{tot}}^{z}e^{-i\hat{H}t}\hat{S}_{\text{tot}}^{z}\rho_{0}\right]
\end{align}
is the total magnetization response function and $\rho_{0}$ is the initial state of the spin system. For typical NMR experiments, it is a good approximation to assume that the system starts in an infinite temperature state  $\rho_{0}=\frac{\hat{I}}{\text{Tr}\left[\hat{I}\right]}$, where $\hat{I}$ is the identity operator.

Letting $\ket{z_{j}}$ be the eigenstates
of $\hat{S}_{\text{tot}}^{z}$ corresponding to eigenvalues $m_{j}$,
the response function is computed as
\begin{equation}\label{eq:ResponseFunc}
S\left(t\right)=2\sum_{j;m_{j}>0}\frac{m_{j}}{2^{N_{s}}}\braket{z_{j}\left(t\right)|\hat{S}_{\text{tot}}^{z}|z_{j}\left(t\right)}
\end{equation}
where $\ket{z_{j}\left(t\right)}=e^{-i\hat{H}t}\ket{z_{j}}$ and $N_{s}$
is the number of spins in the system. The quantum algorithm is
thus preparation of the desired computational basis states $\left\{ \ket{z_{j}}\right\} $,
Hamiltonian simulation of $\hat{H}$ through implementation of the
time-evolution operator $\hat{U}\left(t\right)=e^{-i\hat{H}t}$, and
projective measurements in the computational basis. These measurements yield the response function, $S\left(t\right)$, whose Fourier transform gives the desired spectrum, $A\left(\omega\right)$.

We implement this time-evolution using a first-order Trotter decomposition into gates commonly used in trapped ion platforms. Specifically, we split the total
time-evolution into $r$ Trotter steps yielding $\hat{U}\left(t\right)=\left[\hat{U}\left(\Delta t\right)\right]^{r}$
where $\Delta t=\frac{t}{r}$. The unitary $\hat{U}\left(\Delta t\right)=e^{-i\hat{H}\Delta t}$
is then approximated with the Suzuki-Trotter product formula
\begin{multline}
\hat{U}_{1}\left(\Delta t\right) =e^{-i\left(\sum_{i}h_{i}\hat{S}_{i}^{x}\right)\Delta t}\left(\Pi_{\left\langle ij\right\rangle }e^{-i\hat{S}_{i}^{z}\hat{S}_{j}^{z}\left(2J_{ij}\Delta t\right)}\right)\times\\
\times\left(\Pi_{\left\langle ij\right\rangle }e^{-i\hat{S}_{i}^{y}\hat{S}_{j}^{y}\left(2J_{ij}\Delta t\right)}\right)\left(\Pi_{\left\langle ij\right\rangle }e^{-i\hat{S}_{i}^{x}\hat{S}_{j}^{x}\left(2J_{ij}\Delta t\right)}\right)
\end{multline}
where $\left\langle ij\right\rangle $ corresponds to all unique pairs
of spins as $J_{ij}=J_{ji}$ in the Hamiltonian. Furthermore, we only
include pairs of spins where $J_{ij}\neq0$. The total time-evolution
is then given by $\hat{U}_{1}\left(t\right)=\left[\hat{U}_{1}\left(\Delta t\right)\right]^{r}$.
Defining the two-qubit gates $\hat{U}^{\alpha\alpha}\left(\phi_{ij}\right)= \exp\{-i\hat{S}_{i}^{\alpha}\hat{S}_{j}^{\alpha}\phi_{ij}\}$
where $\phi_{ij}=2J_{ij}\Delta t$, single-qubit rotation gates
$\hat{R}_{i}^{\alpha}\left(\phi\right)=e^{-i\hat{S}_{i}^{\alpha}\frac{\phi}{2}}$,
and angles $\phi_{i}=2h_{i}\Delta t$, the quantum circuit for
time evolution is given by 
\begin{multline}\label{eq:trotFormula}
\hat{U}_{1}\left(t\right) =\Pi_{m=1}^{r}\{
\left(\Pi_{i}\hat{R}_{i}^{y}\left(-\frac{\pi}{2}\right)\right)\left(\Pi_{i}\hat{R}_{i}^{z}\left(-\phi_{i}\right)\right)\times\\
\times\left(\Pi_{\left\langle ij\right\rangle }\hat{U}^{xx}\left(\phi_{ij}\right)\right)\left(\Pi_{i}\hat{R}_{i}^{y}\left(\frac{\pi}{2}\right)\right) \times\\
\times\left(\Pi_{i}\hat{R}_{i}^{z}\left(-\frac{\pi}{2}\right)\right)\left(\Pi_{\left\langle ij\right\rangle }\hat{U}^{xx}\left(\phi_{ij}\right)\right)\times\\
\times\left(\Pi_{i}\hat{R}_{i}^{z}\left(\frac{\pi}{2}\right)\right)
\left(\Pi_{\left\langle ij\right\rangle }\hat{U}^{xx}\left(\phi_{ij}\right)\right)\}
\end{multline}
where we apply gates from right to left. The Trotter decomposition, Eq.~\eqref{eq:trotFormula}, is expressed in terms of the Molmer-Sorensen gates, $\hat{U}^{xx}\left(\phi_{ij}\right)$, and single qubit rotations that are commonly used in trapped ion computations.

\begin{figure}[t!]
	\centering
	\includegraphics[clip, width=0.99\columnwidth]{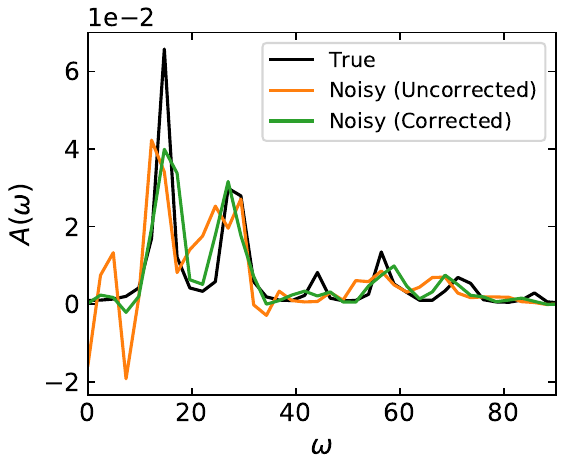}
	\caption{Example spectrum without noise (black), with Trotter error and unitary noise (orange), and noise with feedforward control (green) for a system of four spins evolving under the Heisenberg Hamiltonian, Eq.~\eqref{eq:HeisenbergHam}. The phonon heating rate is taken to be $c_{2}=0.02$ ms$^{-1}$ and the noisy spectra are averaged over 40 runs. The uncorrected noisy spectrum is computed using 200 gates and the corrected noisy spectrum is computed using 500 gates, which are the gate depths for which each spectrum is closest to the noiseless spectrum, as quantified by the Hellinger distance between the spectra.}
	\label{fig:spectrum}
\end{figure}

Assuming that enough measurements are made during a computation to ignore measurement errors in the expection values, $\braket{z_{j}\left(t\right)|\hat{S}_{\text{tot}}^{z}|z_{j}\left(t\right)}$, the computed spectrum will still include discretization errors from Trotterization and unitary gate errors from the ion motion described in Sec.\ref{sec:NoiseModel}. The feedforward control discussed in Sec.\ref{sec:controlFramework} can help mitigate the latter. Figure~\ref{fig:spectrum} shows an example spectrum (black), the same spectrum with both Trotter and unitary noise (orange), and the noisy spectrum with feedforward control (green). 

\section{Trapped Ion Noise Model}
\label{sec:NoiseModel}

Trapped ions have emerged as a leading platform for quantum computation and simulation due to their long coherence times, identical nature, and negligible idle errors~\cite{Ladd2010, Cetina2022}. The ions in these systems crystallize into a chain after being tightly confined in two directions via an oscillating electric field. Entanglement between the  qubits is generated by a laser-induced interaction between states that is mediated by the collective motion of ions. Usually, the motional modes along the tightly confined transverse direction are used for these operations as they are less sensitive to electric field fluctuations arising from the electrodes generating the trap. These fluctuations do, however, excite the weakly confined longitudinal modes of the chain. The deviation of the ions from their lattice positions causes them to experience erroneous intensities from the individually addressed laser beams used to implement different operations. As the longitudinal motion of the ions heats up, these errors build into a dominant form of noise that limits the operational time window of the system~\cite{Cetina2022}. Here, we develop a noise model for errors arising from this longitudinal heating, ignoring other possible sources of error in trapped ion systems that may be more prevalent in different operational regimes of the device.

We first characterize the gate error in the system due to longitudinal movement of the ions in the $x$-direction, depicted in Fig.~\ref{fig:ionSchematic}. The individually addressed single- and two-qubit gates in trapped ion systems are enacted by shining a narrowly focused laser on a single or pair of ion lattice sites respectively. The gates take the form $\hat{U}\left(\phi\right)=\exp\left(-i\phi\hat{A}\right)$, where $\hat{A}$ is either a single spin operator, $\hat{S}_{j}^{\alpha}$, acting on a site $j$, or the bilinear, $\hat{S}_{i}^{x}\hat{S}_{j}^{x}$, acting on a pair of sites. These gates form a sufficient set for universal quantum computation. The phase of the gate is $\phi=\Omega t_{g}$, where $t_{g}$ is the duration of the laser pulse, and $\Omega$ is the Rabi frequency set by the electric field amplitude of the laser. This amplitude typically has a Gaussian spread in the longitudinal direction which carries over to the Rabi frequency: $\Omega\left(x\right)=\Omega_{0}\exp\left(-x^{2}/(2\sigma^2)\right)$ where $\Omega_{0}$ represents the maximum beam intensity and $\sigma$ characterizes the beam width. The collective motion of the ions in the longitudinal direction can be  decomposed in terms of normal modes with frequencies $\omega_{m}$. During application of a gate, these motional oscillations cause the ions to feel a position-dependent Rabi frequency that is less than the desired $\Omega_{0}$. Our goal is to derive the distribution of the erroneous phase $\phi$ that is applied when inputting an angle $\phi_\textrm{in}=\Omega_{0}t_{g}$. In general, this distribution will evolve in time as the longitudinal phonon modes are heated, leading to larger amplitude oscillations. We therefore also seek to determine how the erroneous Rabi frequency, and therefore the phases $\phi(t)$ and $\phi(t')$, are correlated at different times. Temporal correlations over a sufficiently long timescale can limit the fidelity of computations in the system, even after feedforward optimization of individual gates.

Electric field fluctuations from electrodes trapping the ions are primarily responsible for heating the longitudinal phonons~\cite{Brownnutt2015}.  The lowest frequency phonon mode, characterized by ions oscillating in phase at frequency $\omega_{0}$, typically dominates the gate error as the field fluctuations are roughly uniform over the chain~\cite{Cetina2022}. The gate application time, $t_{g}$, is usually much longer than the timescale set by $\omega_{0}$ so we can assume that the effective Rabi frequency, $\overline{\Omega}(t)$, that an ion feels during a gate initiated at time $t$ only depends on the average position of the ion:
\begin{equation}\label{eq:avgRabi}
	\overline{\Omega}(t) = \Omega_{0}\exp\left(-\frac{\overline{x}^{2}(t)}{2\sigma^2}\right)
\end{equation}
where $\overline{x}(t)=\frac{1}{t_{g}}\int_{t}^{t+t_{g}}ds\left\langle\hat{x}(s)\right\rangle$ and $\hat{x}$ is the position operator of the ion. Letting $\hat{p}$ be the canonically conjugate ion momentum operator, we define the usual bosonic creation and annihilation operators $\hat{a}^{\dagger}=\left(\hat{x}-i\hat{p}\right)/\sqrt{2}$ and $\hat{a}=\left(\hat{x}+i\hat{p}\right)/\sqrt{2}$. The average ion position only depends on the average energy of the harmonic motion: $\overline{x}^{2}(t)=\hbar\left\langle\hat{n}(t)\right\rangle/\left(m\omega_{0}\right)$ where $m$ is the mass of the ion and $\hat{n}=\hat{a}^{\dagger}\hat{a}$ is the occupation number. 

We must describe the dynamics of the ions' harmonic motion in order to compute the distribution and correlations of the Rabi frequencies, and by extension the phases of the unitary gate. Letting the state of the system be $\rho\left(t\right)$, we can model the dynamics with the Lindblad master equation
\begin{align}\label{eq:masterEq}
	\frac{d}{dt}\rho &= -\frac{i}{\hbar}\left[\hbar\omega_{0}\hat{n},\rho\right]\nonumber\\
							&+ \gamma_{+}\left(\hat{a}^{\dagger}\rho\hat{a}+\frac{1}{2}\left\{\hat{a}\hat{a}^{\dagger},\rho\right\}\right)\nonumber\\
							&+ \gamma_{-}\left(\hat{a}\rho\hat{a}^{\dagger}+\frac{1}{2}\left\{\hat{a}^{\dagger}\hat{a},\rho\right\}\right),
\end{align}
where the first term represents the coherent harmonic oscillation of the ions, the second term represents an increase in the oscillation amplitude at rate $\gamma_{+}$, and the third term represents a decrease in the oscillation amplitude at rate $\gamma_{-}$. These latter two terms describe the incoherent dynamics of the ions resulting from background electric field fluctuations. Assuming that this background field exists in a thermal state at temperature $T$, the ions' oscillation amplitude changes at rates $\gamma_{+}=\gamma\mathcal{N}\left(\omega_{0},T\right)$ and $\gamma_{-}=\gamma\left(\mathcal{N}\left(\omega_{0},T\right)+1\right)$, where $\mathcal{N}\left(\omega_{0},T\right)=1/\left(e^{\hbar\omega_{0}/k_{B}T}-1\right)$ is the Bose-Einstein distribution of the electric field occupation. We assume that the background electric field is at infinite temperature so both these rates are equal and redefine $\gamma$ such that $\gamma_{+}=\gamma_{-}=\gamma$. Given that the relevant phonon frequencies are of the order of a few hundreds of kHz, this approximation is satisfied down to very low temperatures~\cite{Brownnutt2015}. We also assume that the laser pulse enacting the gate does not affect the ions' motional state; in this sense, it is a weak measurement rather than a strong measurement which would collapse the ions' motion into a particular eigenstate of the occupation $\hat{n}$.

In trapped ion experiments, it is possible to cool the chain close to its motional ground state during preparation of the system. We therefore assume that the initial motional state of the system is the phonon vacuum $\rho\left(t_{0}\right)=\ket{0}\bra{0}$. Dynamics under Eq.~\eqref{eq:masterEq} will then evolve the system into a harmonically oscillating coherent state undergoing a diffusive random walk in its amplitude. It therefore makes sense to describe the system state in terms of its Glauber-Sudarshan P-function representation:
\begin{equation}\label{eq:Pfunc}
	\rho = \int d^{2}\alpha P\left(\alpha,\alpha^{*},t\right)\ket{\alpha}\bra{\alpha}
\end{equation}
where $\{\ket{\alpha}\}$ are coherent states that form a basis for the system. The dynamics of the system is then captured by a Fokker-Planck equation for the P-function, $P\left(\alpha,\alpha^{*},t\right)$,
\begin{equation}\label{eq:FokkerPlanck}
	\frac{d}{dt}P = \left\{ i\omega_{0}\left(\frac{\partial}{\partial\alpha}\alpha - \frac{\partial}{\partial\alpha^{*}}\alpha^{*}\right) +\gamma\frac{\partial^2}{\partial\alpha\partial\alpha^{*}}\right\} P.
\end{equation}

\begin{figure*}[t!]
	\centering
	\includegraphics[width=0.99\textwidth]{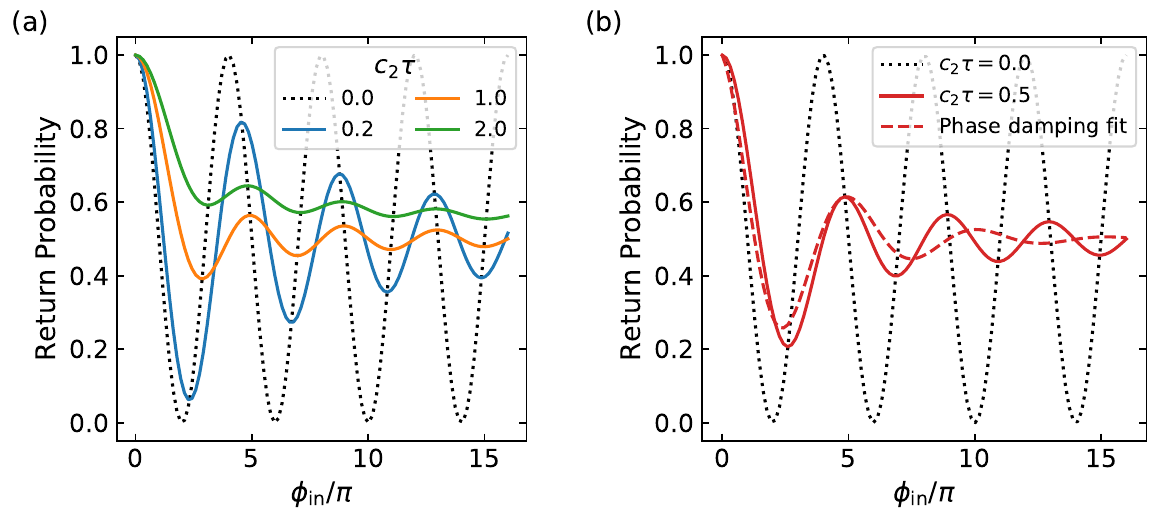}
	\caption{Return probability predictions for experimental protocol to extract $c_{2}$. (a) Curves for different wait times, $\tau$, as a function of input angle $\phi_\textrm{in}$. (b) Difference between derived noise model and phase damping.}
	\label{fig:NoiseEcho}
\end{figure*}

The Green's function of the Fokker-Planck equation, expressed in the rotating frame of the phonon mode with frequency $\omega_0$, is
\begin{equation}\label{eq:FPGreensFunc}
	K\left(\alpha', t' \lvert \alpha,t\right) = \frac{1}{\pi\gamma\left(t'-t\right)}\exp\left\{-\frac{\lvert \alpha'-\alpha\rvert}{\gamma\left(t'-t\right)}\right\},
\end{equation}
which can be interpreted as the probability to find the ions in state $\ket{\alpha'}$ at time $t'$ given that they were in state $\ket{\alpha}$ at time $t$. To compute the full state of the system one simply has to involve this kernel with the initial P-function. For systems that have not been fully cooled down and maintain a non-zero thermal occupation of phonons, the initial state will be Gaussian. At present, it suffices to consider the $T=0$ limit in which case the P-function is simply $P(\alpha,\alpha^\ast)=\delta(\alpha)\delta(\alpha^\ast)$.
This Green's function can be used to compute the probability distribution and correlations of observables expressed in the coherent state basis. Letting $\phi\left(\tau\right)=\overline{\Omega}\left(\tau\right)t_{g}$ be the angle imparted by a unitary gate applied at time $\tau$ in the experiment, when the phonon mode is in state $\ket{\alpha\left(\tau\right)}$, the Rabi phase of the qubit will advance by an angle
\begin{equation}\label{eq:avgPhi}
	\phi\left(\tau\right)=\phi_\textrm{in}\exp\left\{-\left(\frac{a_\textrm{osc}}{a_\textrm{laser}}\right)^{2}\lvert\alpha\left(\tau\right)\rvert^{2}\right\},
\end{equation}
where $a_\textrm{osc}=\sqrt{\hbar/(m\omega_{0})}$ and $a_\textrm{laser}=\sqrt{2}\sigma$ are the characteristic length scales of the harmonic oscillator and Gaussian laser respectively. The probability distribution of the angle can then be computed as
\begin{equation}\label{eq:phiProbDist}
	p_{\phi}\left(\phi;\tau,c_{2}\right)=\frac{1}{c_{2}\tau}\frac{1}{\phi}\left(\frac{\phi}{\phi_\textrm{in}}\right)^{\frac{1}{c_{2}\tau}}\Theta\left(\phi_\textrm{in}-\phi\right),
\end{equation}
where we have defined the heating rate constant 
\begin{equation}\label{eq:c2}
	c_{2}=\gamma\left(\frac{a_\textrm{osc}}{a_\textrm{laser}}\right)^{2}
\end{equation}
and the Heaviside step function, $\Theta\left(\phi_\textrm{in}-\phi\right)$, encodes the fact that the time-averaged Rabi frequency felt by the ion cannot be more than spending all its time at the center of the laser where its intensity is strongest. The distribution of angles, Eq.~\eqref{eq:phiProbDist}, is the noise model we need for feedforward control. Note that it only depends on a single latent variable, $\lambda=c_{2}\tau$, representing the amount of diffusion the ions' motion has undergone.

We can gain insight into the angle distribution by examining the average and typical angles that are applied by the gate,
\begin{align}
	\phi_\textrm{avg}&=\mathbb{E}_{\phi}\left[\phi\right]=\frac{\phi_\textrm{in}}{1+\lambda}\label{eq:avgAngle}\\
	\phi_\textrm{typ}&=\exp{\left(\mathbb{E}_{\phi}\left[\log\phi\right]\right)}=\phi_\textrm{in}e^{-\lambda}.
\end{align}
We see that at late experimental times compared to the rate $c_{2}$ such that $\lambda\rightarrow\infty$, both the average and typical angles go to zero. Physically, the amplitude of the ions' oscillation becomes so large that the ion never spends time inside the laser beam and hence its internal qubit state is not changed. While the average angle algebraically decays to zero at late times, the typical angle becomes very small as $\tau$ crosses $1/c_{2}$, thus showing that $c_{2}$ sets the timescale where we can coherently manipulate the qubits in an experiment.

We can further understand the effects of noise on a quantum computation or simulation by examining the correlation between two gates applied at a time $\Delta_{\tau}$ apart,
\begin{multline}\label{eq:angleCorr}
	\textrm{Corr}\left(\phi(\tau+\Delta_{\tau})\phi(\tau)\right)=\frac{\textrm{Cov}\left(\phi(\tau+\Delta_{\tau})\phi(\tau)\right)}{\sqrt{\textrm{Var}\left(\phi(\tau)\right)\textrm{Var}\left(\phi(\tau)\right)}}\\\\
	=\frac{\tau}{\tau+\Delta_{t}}\frac{\sqrt{\left(1+2c_{2}\tau\right)\left(1+2c_{2}(\tau+\Delta_{\tau})\right)}}{1+2c_{2}\tau+c_{2}\Delta_{\tau}+c_{2}^{2}\tau\Delta_{\tau}}
\end{multline}
Taking the limit at late times, we have
\begin{equation}\label{eq:angleCorrLateTimes}
	\lim\limits_{\tau\rightarrow\infty}\textrm{Corr}\left(\phi(\tau+\Delta_{\tau})\phi(\tau)\right)=\frac{1}{1+\frac{1}{2}c_{2}\Delta_{\tau}}+\mathcal{O}\left(\frac{1}{\tau}\right),
\end{equation}
which shows that $c_{2}$ also sets the temporal correlation length between different gates. Given that the gate application time, $t_{g}$, is small compared to typical values of $c_{2}$ in trapped ion experiments, the unitary gate errors will be temporally correlated.

As a limiting case, we can examine how the noisy gate angles are distributed at short times when the ions are very close to the center of the laser beam. By simultaneously taking the limits $\phi\rightarrow\phi_\textrm{in}$ and $c_{2}\tau\rightarrow0$ in Eq.~\eqref{eq:phiProbDist}, we get the short time distribution
\begin{equation}\label{eq:phiProbDist_shortTime}
	p_{\phi}^\textrm{short}\left(\phi;\tau,c_{2}\right)=\frac{1}{c_{2}\tau\phi_\textrm{in}}e^{-\frac{\left(\phi_\textrm{in}-\phi\right)}{c_{2}\tau\phi_\textrm{in}}}\Theta\left(\phi_\textrm{in}-\phi\right).
\end{equation}
This expression can equivalently be derived by Taylor expanding Eq.~\eqref{eq:avgPhi} as $\phi\left(\tau\right)=\phi_\textrm{in}\left(1-(\frac{a_\textrm{osc}}{a_\textrm{laser}})^{2}\lvert\alpha\left(\tau\right)\rvert^{2}\right)$ and computing the probability distribution of gate angles using the Green's function given in Eq.~\eqref{eq:FPGreensFunc}. The exponential distribution of gate angles described in Eq.~\eqref{eq:phiProbDist_shortTime}, valid at short times, is in agreement with the ion noise model discussed in Ref.~\cite{Cetina2022}.

We now give a protocol to experimentally extract the value of $c_{2}$ which characterizes the noise in a particular trapped ion set-up. Prepare a system of two qubits in the computational basis state $\ket{\downarrow\downarrow}$, wait a time $\tau$, and apply a gate
$\hat{U}_{xx}\left(\phi\right)=\exp\left(-i\phi\hat{S}_{i}^{x}\hat{S}_{j}^{x}\right)$ with an input angle $\phi_\textrm{in}$. Then, do a projective measurement in the computational basis state to extract the return probability of the system being in the $\ket{\downarrow\downarrow}$ state. If there was no noise in the system, this probability would be
\begin{equation}
	P_{\downarrow\downarrow}=\braket{\downarrow\downarrow\lvert\hat{U}_{xx}\left(\phi_\textrm{in}\right)\rvert\downarrow\downarrow}=\cos^{2}\left(\frac{\phi_\textrm{in}}{4}\right)
\end{equation} for all $\tau$. With unitary gate error due to the ions' motion, the probability becomes
\begin{multline}\label{eq:avgReturnProb}
	\overline{P}_{\downarrow\downarrow}\left(\phi_\textrm{in},c_{2}\tau\right) = \mathbb{E}_{\phi}\left[\cos^{2}\left(\frac{\phi}{4}\right)\right]\\
	=   \cos^{2}\left(\frac{\phi_\textrm{in}}{4}\right)
	+ \frac{\phi_\textrm{in}^{2}c_{2}\tau}{8+16c_{2}\tau}\times\\
	 \times {}_{1}F_{2} (1 + \frac{1}{2c_{2}\tau}, \frac{3}{2}, 2 + \frac{1}{2 c_{2}\tau}, -\frac{\phi_\textrm{in}^{2}}{16}),	
\end{multline}
where ${}_{1}F_{2}$ is the generalized hypergeometric function. This average return probability is directly related to the moment generating function of Eq.~\eqref{eq:phiProbDist}. Measuring Eq.~\eqref{eq:avgReturnProb} for different input angles, $\phi_\textrm{in}$, and wait times, $\tau$, yield curves that can be used to fit the value $c_{2}$. We give examples of these curves in Fig.~\ref{fig:NoiseEcho}(a). In Fig.~\ref{fig:NoiseEcho}(b), we show how the return probability can differentiate between the noise model derived here and and typical phase damping. The latter leads to a return probability characterized by an exponentially decaying oscillations with a constant phase shift dependent on the input angle. Armed with knowledge of the noise model, Eq~\eqref{eq:phiProbDist}, and a method to experimentally determine the latent variable, $c_{2}$, we now illustrate how non-Markovian correlations in the noise induce an optimal gate depth when implementing a quantum algorithm.

\begin{figure*}[t!]
	\centering
	\includegraphics[width=0.99\textwidth]{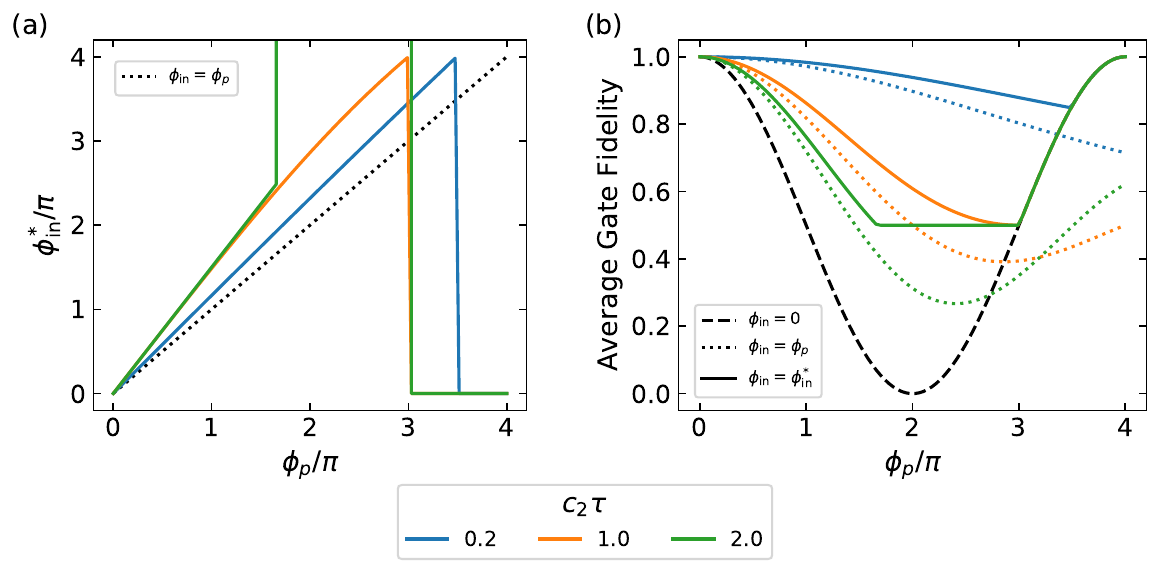}
	\caption{Optimal feedforward control characterization. (a) Optimal input angle. (b) Average gate fidelity. The black dashed line depicts the fidelity if no gate is applied and the dotted lines represent the fidelity if the desired output angle, $\phi_{p}$, is directly taken as the input to the gate.}
	\label{fig:optimalFeedForward}
\end{figure*}

\section{Optimal gate depth}\label{sec:optimalGateDepth}

We can gain insight into how non-Markovian correlations amongst gates induce an optimal gate depth in a quantum algorithm by first considering a single one- or two-qubit gate $\hat{U}\left(\phi_\textrm{tot}\right)=\exp\left(-i\phi_\textrm{tot}\hat{A}\right)$ of the form discussed in Sec~\ref{sec:NoiseModel}. Let us discretize this gate into $r$ Trotter steps: $\hat{U}\left(\phi_\textrm{tot}\right)=\left[\hat{U}\left(\phi\right)\right]^{r}$ where $\phi=\phi_\textrm{tot}/r$ and  $\hat{U}\left(\phi\right)=\exp\left(-i\phi\hat{A}\right)$. The expected angle applied by the total sequence $\hat{U}\left(\phi_\textrm{tot}\right)$ is
\begin{equation}
	\mathbb{E}\left[\phi_\textrm{tot}\right] = r\mathbb{E}_{\phi}\left[\phi\right]
\end{equation}
where $\mathbb{E}_{\phi}\left[\phi\right]$ is the average angle applied by $\hat{U}\left(\phi\right)$. If the unitary gate errors in the system were modeled as Markovian, and therefore uncorrelated, the variance in the total angle would be
\begin{equation}
	\textrm{Var}\left(\phi_\textrm{tot}\right) = r\textrm{Var}\left(\phi\right)
\end{equation}
where $\textrm{Var}\left(\phi\right)$ is the variance in the angle applied by $\hat{U}\left(\phi\right)$. Regardless of the source of unitary error, this variance of each discretized gate will typically be proportional to $\mathbb{E}_{\phi}\left[\phi\right]^2$. Letting the constant of proportionality be $\beta$, defined through $\textrm{Var}\left(\phi\right)=\beta\mathbb{E}_{\phi}\left[\phi\right]^2$, the noise-to-signal ratio of the total gate sequence becomes:
\begin{equation}
	\eta=\frac{\sqrt{\textrm{Var}\left(\phi_\textrm{tot}\right)}}{\mathbb{E}\left[\phi_\textrm{tot}\right]}=\frac{\sqrt{\beta}}{r}.
\end{equation}
As an example, if we take the noise model developed in Sec.~\ref{sec:NoiseModel} and ignore temporal correlations, we have $\beta=\left(c_{2}\tau\right)^{2}/(1+2c_{2}\tau)$. This constant $\beta$ is computed by assuming that each gate angle is independent and identically distributed according to Eq.~\eqref{eq:phiProbDist}. We see that $\eta\rightarrow0$ as $r\rightarrow\infty$, implying that discretizing the total intended gate, $\hat{U}\left(\phi_\textrm{tot}\right)$, into a large number of steps eliminates the unitary error in the system, thus illustrating that unitary gate errors in experiments cannot be fully described using a Markovian noise model. Correlations between the unitary gate errors are responsible for decoherence observed in experiments, with an optimal gate depth being set by the timescale upon which this decoherence becomes too large.


To demonstrate how this optimal gate depth manifests in practice, we turn to the Hamiltonian simulation task described in Sec.~\ref{sec:HamSim}. The computed spectrum will have errors both due to discretization via the Trotter decomposition, Eq.~\eqref{eq:trotFormula}, and unitary gate noise due to heating of the ions' motion as described in Sec.~\ref{sec:NoiseModel}. Trotter error decreases as the number of gates in the circuit is increased, while unitary errors accumulate as the number of gates is increased. Therefore, there is an optimal gate count balancing Trotter error and accumulated unitary error. 

We can quantify the error in the computation use two different metrics. The first is to compute the average fidelity
\begin{equation}\label{eq:totFidelity}
F\left(t\right)=\lvert\frac{1}{2^{n}}\textrm{Tr}\{\hat{U}_{1}\left(t\right)^{\dagger}\hat{U}\left(t\right)\}\rvert^2,
\end{equation}
where $\hat{U}\left(t\right)=e^{-i\hat{H}t}$, with $\hat{H}$ given by Eq.~\eqref{eq:HeisenbergHam}, is the desired time evolution operator and $\hat{U}_{1}\left(t\right)$ is the noisy Trotterized evolution we implement in the quantum circuit, given by Eq~\eqref{eq:trotFormula}, with noisy gate angles. Given that computation of a spectrum requires implementing time-evolution for a series of different times in order to generate samples of $S(t)$, given by Eq.~\eqref{eq:ResponseFunc}, we can define the time-integrated fidelity
\begin{equation}\label{eq:intFidelity}
	F_\textrm{int} = \frac{1}{T}\int_{0}^{T}F\left(t\right)
\end{equation} 
where $T$ is the last sampled time. The optimal gate depth is then determined by the largest value of $F_\textrm{int}$. This metric is not biased towards any particular choice of observable.

Alternatively, we can quantify the error in the spectrum by computing the Hellinger distance
\begin{equation}\label{eq:HellingerDist}
D_{H}^{2}\left(A_{i},A_{j}\right)=\frac{1}{2}\int\frac{d\omega}{2\pi}\left(\sqrt{A_{i}(\omega)}-\sqrt{A_{j}(\omega)}\right)^{2}
\end{equation}
between a noiseless spectrum, $A_{i}(\omega)$, generated by the perfect time evolution operator, $\hat{U}\left(t\right)=e^{-i\hat{H}t}$, and a noisy spectrum, $A_{j}(\omega)$, generated by a noisy Trotterized evolution, $\hat{U}_{1}\left(t\right)$ . At the optimal gate depth, the Trotterized spectrum will have the most overlap with the true noiseless spectrum according to the Hellinger distance. This metric is biased towards the computation of the spectrum.

\begin{figure*}[t!]
	\centering
	\includegraphics[width=0.9\textwidth]{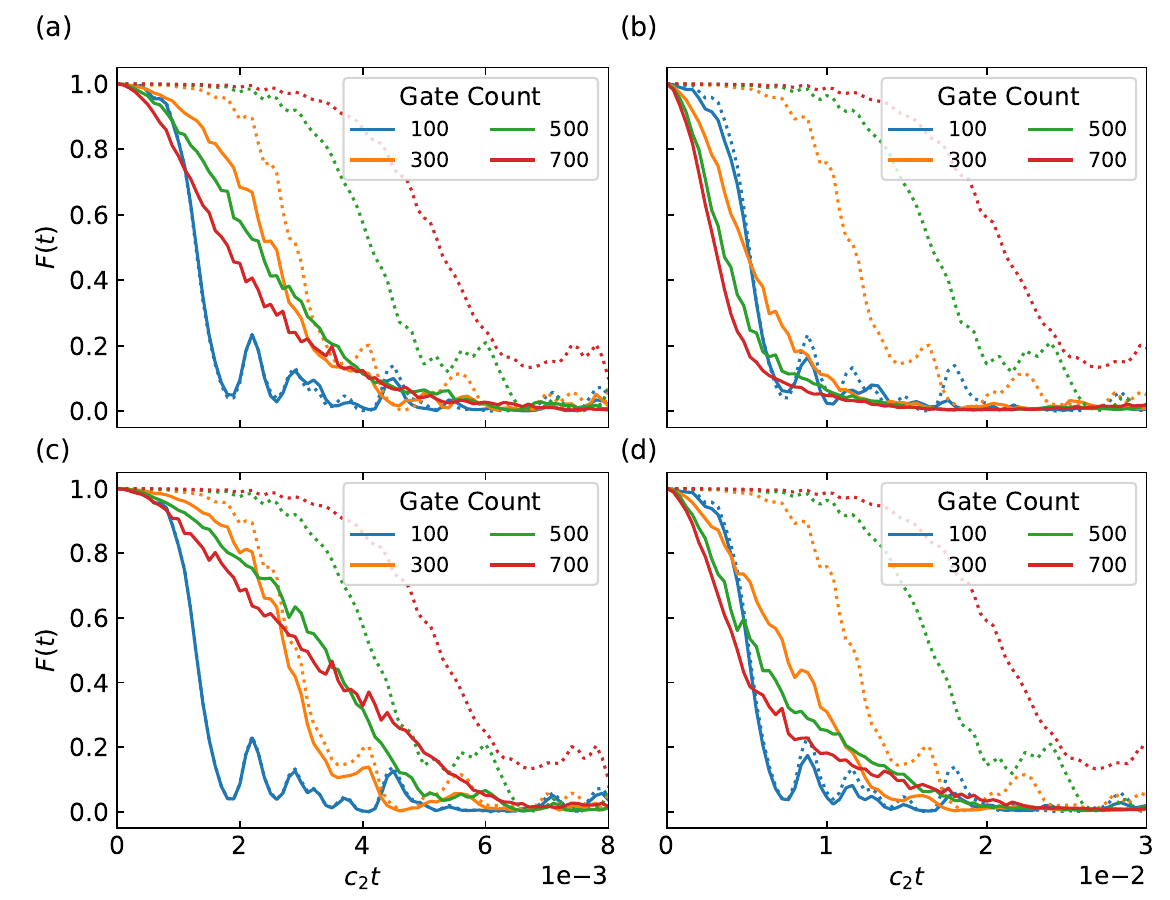}
	\caption{Time dependent fidelity of noisy Trotterized time evolution for a system of four spins evolving under the Heisenberg Hamiltonian, Eq.~\eqref{eq:HeisenbergHam}. Solid curves include both heating noise and Trotter error, while the dotted curves include only Trotter error and are given as a noiseless reference. (a) $c_{2}=0.005$ ms$^{-1}$ and no feedforward correction. (b) $c_{2}=0.02$ ms$^{-1}$ and no feedforward correction. (c) $c_{2}=0.005$ ms$^{-1}$ with feedforward correction. (d) $c_{2}=0.02$ ms$^{-1}$ with feedforward correction. The noisy computations are averaged over 40 runs.}
	\label{fig:fidTimeTrace}
\end{figure*}

The optimal gate depth with the corresponding average fidelity and Hellinger distance for an example noisy computation is shown in Fig.~\ref{fig:optimalFid} and Fig.~\ref{fig:optimalSpectrum} respectively, and we discuss these results in the next section. The total amount of error in the noisy computation can be reduced by appropriately modifying the angles of the gates comprising the quantum simulation circuit, Eq.~\eqref{eq:trotFormula}, a method known as feedforward control. We develop a systematic, platform-independent protocol to determine the modified gate angles in the next section. We then illustrate the benefits of the feedforward control in the context of the Hamiltonian simulation task by showing improvements in the fidelity and Hellinger distance for an example Hamiltonian of the form in Eq.~\eqref{eq:HeisenbergHam}.

\section{Feedforward Control}
\label{sec:controlFramework}

A quantum computation or simulation involves applying a unitary operation $\hat{U}$ to a system of qubits. Often, this unitary transformation is a composite of several single- and two-qubit unitary gates $\hat{U}\left(\phi\right)=\exp\left(-i\phi\hat{A}\right)$, with $\hat{A}$ typically linear or bilinear in spin-$1/2$ operators, $\hat{S}_{j}^{\alpha}=\hat{\sigma}_{j}^{\alpha}/2$. A unitary error in the system manifests as application of $\hat{U}\left(\phi\right)$ when we intend to apply $\hat{U}\left(\phi_{p}\right)$. We usually do not have deterministic knowledge of the value of the incorrect angle, $\phi$, and therefore describe it with a probability distribution $p_{\phi}\left(\phi;\phi_\textrm{in},\vec{\lambda}\right)$, where $\vec{\lambda}$ is a vector of latent variables characterizing the physical noise underlying the system and $\phi_\textrm{in}$ is the angle we input when applying the gate. If the gate was noiseless, we would have $p_{\phi}\left(\phi;\phi_\textrm{in},\vec{\lambda}\right)=\delta\left(\phi-\phi_\textrm{in}\right)$ and would input $\phi_\textrm{in}=\phi_{p}$, where $\phi_{p}$ is the desired output gate angle. The idea of feedforward control is to appropriately adjust the input gate angles of the computation to reduce the error accumulated from incorrect gate angles.

Formally, let the total unitary describing the actual computation be $\hat{U}=\prod_{m=1}^{M}\hat{U}_{m}\left(\phi^{(m)}\right)$, where  $\hat{U}_{m}\left(\phi^{(m)}\right)=\exp\left(-i\phi^{(m)}\hat{A}_{m}\right)$. The desired computation is $\hat{U}_{p}=\prod_{m=1}^{M}\hat{U}_{m}\left(\phi_{p}^{(m)}\right)$.  As the output angles are probabilistic, a particular manifestation of the output computation $\hat{U}$ depends on the joint probability distribution $p_{\vec{\phi}}\left(\vec{\phi};\vec{\phi}_\textrm{in},\vec{\lambda}\right)$, where $\vec{\phi}_\textrm{in}$ and $\vec{\phi}$ are the $m$ different input and output angles respectively. The goal of feedforward control is to pick the optimal input angles, $\vec{\phi}_\textrm{in}^{*}$, such that the computation $\hat{U}$ is close to $\hat{U}_{p}$ on average. In general, $\vec{\phi}_\textrm{in}^{*}$ will depend on both the set of desired output angles, $\vec{\phi}_{p}$, and the latent noise variables, $\vec{\lambda}$. 

Optimizing over the entire computation, however, can be challenging as it requires knowledge of the full joint distribution, $p_{\vec{\phi}}$, which is generally non-trivial to compute, even for the model presented in Sec.~\ref{sec:NoiseModel}. Additionally, even if possible, such an optimization may not generalize well to other computations represented by different gate sequences. We therefore focus on optimizing each individual unitary gate independently of the others, which amounts to neglecting correlations between unitary gate errors and assuming that they are independent and identically distributed according to the marginal distribution, $p_{\phi}$. Mathematically, this amounts to the factorization of the joint distribution: $p_{\vec{\phi}}=\prod_{m=1}^{M}p_{\phi^{(m)}}$. Temporal correlations in the physical noise underlying the system lead to correlations in the angles $\vec{\phi}$ that are not captured by such a factorization. Feedforward optimization of individual gates can therefore only partially mitigate the error in the overall computation. Ignoring the non-Markovian effects discussed in Sec.~\ref{sec:optimalGateDepth} implies that the feedforward control is best applied at sufficiently shallow gate depths; when circuit discretization becomes comparable to temporal correlations of noise, the feedforward correction will be inaccurate. The advantage of ignoring error correlations, however, is that the correction can be easily applied to any computation, $\hat{U}$, as it done at the level of individual gates.

The error due to applying a gate $\hat{U}\left(\phi\right)$ when we desire to apply $\hat{U}\left(\phi_{p}\right)$ can be quantified by the gate fidelity
\begin{equation}\label{eq:HaarFidelity}
F\left(\phi,\phi_{p}\right)=\lvert\frac{1}{2^{n}}\textrm{Tr}\{\hat{U}\left(\phi\right)^{\dagger}\hat{U}\left(\phi_{p}\right)\}\rvert^2,
\end{equation}
which describes the expected fidelity of an  $n$-qubit gate for a random state drawn uniformly from the $n$-qubit state space~\cite{Nielsen2002}. For example, let us consider a unitary gate corresponding to $\hat{A}=\hat{S}_{i}^{\alpha}\hat{S}_{j}^{\alpha}$ describing an interaction between two qubits $i$ and $j$. The fidelity then takes the simple form $F\left(\phi,\phi_{p}\right)=\cos^{2}\left((\phi-\phi_{p})/4\right)$. The figure of merit we want to optimize with feedforward control is the average fidelity over all possible wrong angles $\phi$,
\begin{align}\label{eq:avgFidelity}
\mathcal{F}\left(\phi_\textrm{in},\phi_{p},\vec{\lambda}\right) = \int d\phi p_{\phi}\left(\phi;\phi_\textrm{in},\vec{\lambda}\right) F\left(\phi,\phi_{p}\right).
\end{align}
The optimal input angle is then 
\begin{equation}\label{eq:optInputAngle}
\phi_\textrm{in}^{*}\left(\phi_{p},\vec{\lambda}\right)=\operatorname*{arg\,max}_{\phi_\textrm{in}} \mathcal{F}\left(\phi_\textrm{in},\phi_{p},\vec{\lambda}\right)
\end{equation}
Calculation of this optimal feedforward angle requires knowledge of the control landscape defined by the dependence of the figure of merit, Eq.~\eqref{eq:avgFidelity}, on the input angle $\phi_\textrm{in}$ and desired output angle $\phi_{p}$.  This landscape can either be numerically mapped out with experimental measurements, or analytically computed after developing a theoretical description of the noise underlying the system.

\begin{figure*}[t!]
	\centering
	\includegraphics[width=0.99\textwidth]{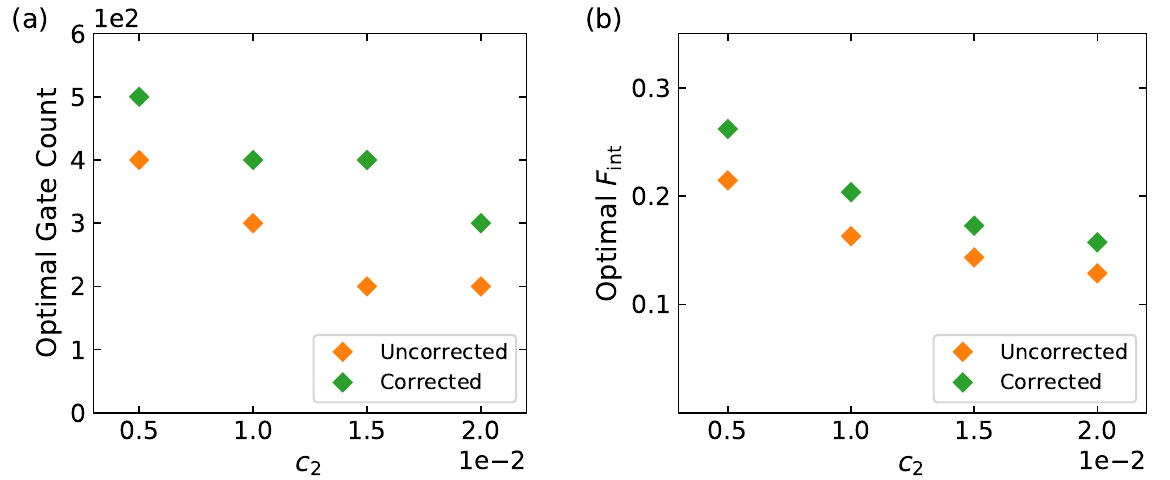}
	\caption{Optimal fidelity resulting from balancing Trotter and decoherence errors for a system of four spins evolving under the Heisenberg Hamiltonian, Eq.~\eqref{eq:HeisenbergHam}. The heating rate $c_{2}$ is given in units of ms$^{-1}$. (a) Optimal gate count. (b) Integrated fidelity. The noisy computations are averaged over 10 runs.}
	\label{fig:optimalFid}
\end{figure*}

As an example of the latter approach, the distribution $p_{\phi}\left(\phi;\phi_\textrm{in},\vec{\lambda}\right)$  for the trapped ion noise discussed in Sec~\ref{sec:NoiseModel} is given by  Eq.~\eqref{eq:phiProbDist}. The ion noise is parameterized by a single latent variable, $\lambda=c_{2}\tau$, which can be experimentally extracted by measuring the return probability Eq.~\eqref{eq:avgReturnProb}. The figure of merit for feedforward control, Eq.~\eqref{eq:avgFidelity}, in this case can be analytically computed:
\begin{multline}\label{eq:ionAvgFidelity}
	\mathcal{F}\left(\phi_\textrm{in},\phi_{p},c_{2}\tau\right) = \frac{1}{2} + \frac{1}{2}\cos\left(\frac{\phi_\textrm{in}}{2}\right)\cos\left(\frac{\phi_{p}}{2}\right)+\\
	+\frac{\phi_\textrm{in}^{2}c_{2}\tau}{8+16c_{2}\tau}\cos\left(\frac{\phi_{p}}{2}\right) {}_{1}F_{2}\left(1+\frac{1}{2c_{2}\tau}, \frac{3}{2},2+\frac{1}{2c_{2}\tau},-\frac{\phi_\textrm{in}^{2}}{16}\right)+\\
	+\frac{\phi_\textrm{in}}{4+4c_{2}\tau}\sin\left(\frac{\phi_{p}}{2}\right) {}_{1}F_{2}\left(\frac{1}{2}+\frac{1}{2c_{2}\tau}, \frac{3}{2}, \frac{3}{2}+\frac{1}{2c_{2}\tau}, -\frac{\phi_\textrm{in}^{2}}{16}\right).
\end{multline}
The optimal input angle, $\phi_\textrm{in}^{*}\left(\phi_{p},c_{2}\tau\right)$, for the trapped ion noise is the angle which satisfies the condition
\begin{equation}\label{eq:ionOptAngle}
	\mathcal{F}\left(\phi_\textrm{in}^{*},\phi_{p},c_{2}\tau\right)=F\left(\phi_\textrm{in}^{*},\phi_{p}\right),
\end{equation} 
where we recall that $F\left(\phi,\phi_{p}\right)=\cos\left((\phi-\phi_{p})/4\right)$ is the fidelity of a gate imparting angle $\phi$ when we desire to apply $\phi_{p}$. We implement feedforward control by taking each desired output gate angle, $\phi_{ij}$, of the $\hat{U}^{xx}$ gates in Eq.~\eqref{eq:trotFormula} as $\phi_{p}$ at the experimental time $\tau$ that the gate is applied. The optimality condition, Eq.~\eqref{eq:ionOptAngle}, is then solved numerically for each such gate and the angle $\phi_\textrm{in}^{*}$ is input into the noisy gate rather than $\phi_{ij}$. 

We show the optimal input angle, Fig~\ref{fig:optimalFeedForward}(a), and average gate fidelity, Fig~\ref{fig:optimalFeedForward}(b), for a range of desired output angles $\phi_{p}$. First, we note that the optimal feedforward angle, $\phi_\textrm{in}^{*}$, always yields a better average fidelity than inputting $\phi_{p}$. We see that for small output angles, there is always a finite optimal input angle. For sufficiently large output angles, however, the optimal input angle is $\phi_\textrm{in}^{*}=0$, meaning we do not apply the gate. These angles are such that doing nothing leads to a better fidelity than any non-zero gate we apply. Furthermore, for times $\tau>1/c_{2}$, meaning that the ions' collective motion has undergone a considerable amount of diffusion, there is an intermediate range of angles where the optimal thing to do is apply a maximally strong laser pulse to make $\phi_\textrm{in}^{*}$ as large as possible. In this case, the gate essentially applies a random phase to the state and yields an average fidelity of $1/2$.

\begin{figure*}[t!]
	\centering
	\includegraphics[width=0.99\textwidth]{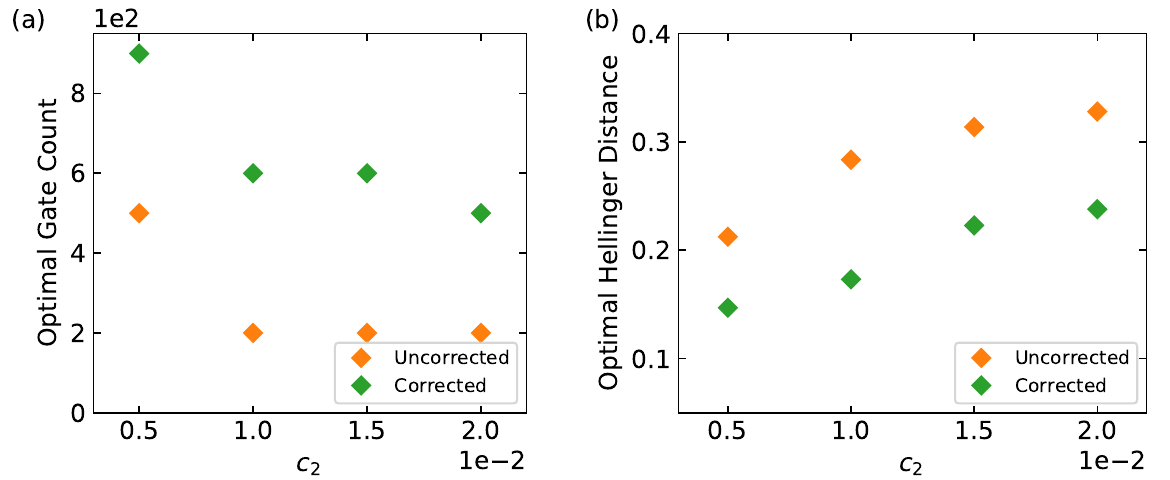}
	\caption{Optimal spectra resulting from balancing Trotter and decoherence errors for a system of four spins evolving under the Heisenberg Hamiltonian, Eq.~\eqref{eq:HeisenbergHam}. The heating rate $c_{2}$ is given in units of ms$^{-1}$. (a) Optimal gate count. (b) Hellinger distance between optimal noisy spectra and noiseless spectrum. The noisy computations are averaged over 10 runs.}
	\label{fig:optimalSpectrum}
\end{figure*}

\begin{figure*}[t!]
	\centering
	\includegraphics[width=0.99\textwidth]{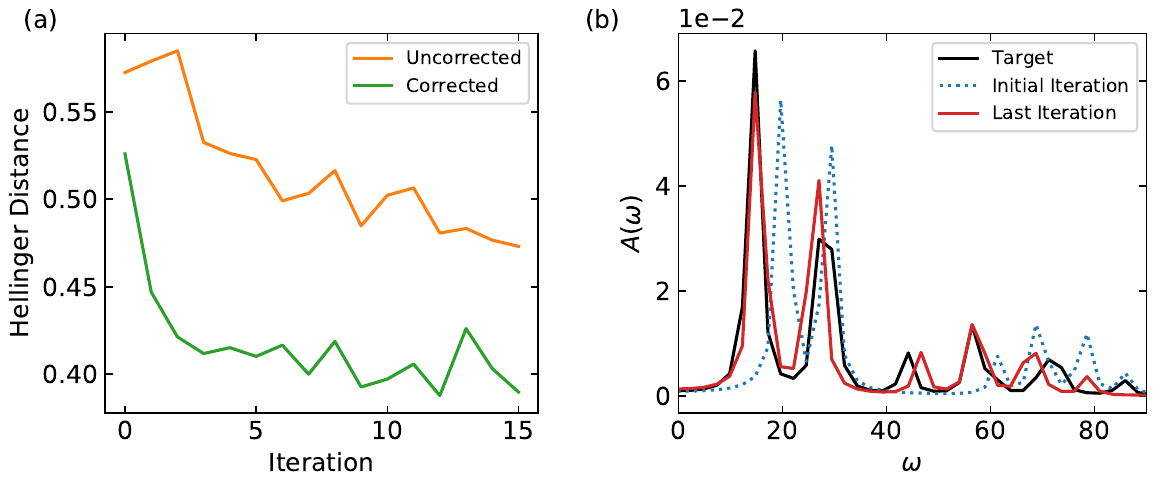}
	\caption{Simulation of NMR inference algorithm with motional noise and Trotter error for a system of four spins. At each update step of the protocol, noisy spectra computed from a set of sample Hamiltonians are used to calculate the next update step (a) Hellinger distance between the average Hamiltonian's noisy spectrum and the target spectrum with and without feedforward correction. (b) Spectrum comparison. We take the average Hamiltonians found at the initial and last iterations of the noisy inference protocol and simulate what its spectrum would be if there was no noise. The fact that the last spectrum is significantly closer to the target spectrum compared to the initial spectrum gives a visual indication of the improvement in the underlying Hamiltonian during the inference protocol. The phonon heating rate is taken to be $c_{2}=0.02$ ms$^{-1}$ and the noisy spectra are computed with 500 gates and averaged over 10 runs. }
	\label{fig:bayesianInf}
\end{figure*}

To benchmark the utility of the feedforward control, we implement with Hamiltonian simulation task of Sec~\ref{sec:HamSim} using optimal input angles computed from Eq.~\eqref{eq:ionOptAngle}. In Fig.~\ref{fig:fidTimeTrace},we plot an example of the time-dependent fidelity, Eq.~\eqref{eq:totFidelity}, for different heating rate and gate counts with and without feedforward control. We see that for low gate counts such as computations with 100 gates, the drop in fidelity comes almost fully from Trotter error without heating noise having much of an effect. For larger gate counts, heating noise becomes the dominant cause of the drop in fidelity. While the fidelity for the zero heating case gets continuously better with increased gate count, finite heating causes computations with sufficiently large gate counts to decrease the overall fidelity. 

Feedforward control can improve the situation in two different ways, which can be seen by comparing, for example, the 300 gate and 700 gate curves. The first effect is to improve the total fidelity over all time values, as quantified by the improvement in $F_\textrm{int}$, Eq.\eqref{eq:intFidelity}, depicted in Fig.~\ref{fig:optimalFid}. This improvement indicate that the computation of $\hat{U}_{1}(t)$ is closer on average to the desired computation $\hat{U}(t)$ for all values of $t$, with the feedforward correction bringing the fidelity of a computation closer to the upper bound set by the Trotter error. The second effect is that for computations with large gate counts, the fidelity for samples at late times, corresponding to large values of $t$, is improved more significantly than for short time samples. This improvement causes the fidelity to have a more shallow decay, and creates windows of time samples where it may be more advantageous to use circuits with different gate counts. For example, in Fig.~\ref{fig:fidTimeTrace}(c), a computation with 300 gates is advantageous for samples with $c_{2}t\lesssim3$, while a computation with 700 gates is advantageous for samples with $c_{2}t\gtrsim3$. The significance of this result is that a particular observables of interest may have information that is more concentrated in a particular time window. For example, the resolution between peak of the spectrum, Eq.~\eqref{eq:Spectrum}, comes from samples at late times. Therefore, the optimal gate count determined by the accuracy of the spectrum may be larger than the optimal gate count determined by the integrated fidelity. Indeed, this is what is seen when comparing Fig.~\ref{fig:optimalFid} and Fig.~\ref{fig:optimalSpectrum}.


 In Fig.~\ref{fig:optimalSpectrum}(a) and (b), we show the optimal gate count and associated Hellinger distance of spectra computed both with and without the feedforward correction. We see that the accuracy of the optimal noisy spectrum is significantly improved. An example spectrum for a system with heating rate $c_{2}=0.02$ ms$^{-1}$ is depicted in Fig.~\ref{fig:spectrum}. The feedforward control both directly mitigates decoherence error from the motion of the ions and indirectly reduces the Trotter error by increasing the optimal gate count. Therefore, by effectively increasing the optimal gate depth of the circuit, feedforward control can be used to partially mitigate both discretization error and accumulated unitary gate error in the system.

\section{Discussion \& Conclusion}
\label{sec:Discussion}

This improvement in the quality of the Hamiltonian simulation can be helpful for practical applications, such as the NMR spectrum inference task discussed in Ref.~\cite{Sels2020}. In that work, a hybrid quantum-classical algorithm is used to infer the parameters of a Hamiltonian, Eq.~\eqref{eq:HeisenbergHam}, that models the system of nuclear spins which produce a given experimental NMR spectrum. The premise of the algorithm is to iteratively simulate the spectrum corresponding to different Hamiltonian parameters on quantum hardware and guess parameters that are closer to the target experimental spectrum using classical optimization techniques. After a sufficient number of iterations, the learned Hamiltonian parameters can be used to gain insight into the chemical structure of the sample that produced the given NMR spectrum. In Fig.~\ref{fig:bayesianInf}, we demonstrate the benefit of feedforward correction in this inference algorithm. Figure~\ref{fig:bayesianInf}(a) shows the Hellinger distance between the average noisy Trotterized spectrum  and a given target spectrum at each iteration of the protocol. We see that the feedforward correction allows the algorithm to converge faster, as the increased resolution in the simulated spectra allows the classical optimization to more easily guess better Hamiltonian parameters. In Fig.~\ref{fig:bayesianInf}(b), we take the Hamiltonian parameters for the initial and last iterations of the noisy protocol with feedforward correction and compute the corresponding spectra without noise to compare how well the learned parameters correspond to the true parameters underlying the given target spectrum. We see that even though the quantum simulation is noisy, we are still able to iteratively infer the Hamiltonian parameters underlying the target spectrum.


We have shown how to tailor gate-based quantum simulation algorithms for particular hardware platforms. Specifically, we demonstrate how knowledge of hardware noise leading to unitary gate errors can be exploited to implement feedforward control to improve the simulation outcome. The ion noise model we derive applies to an array of computations and simulations performed in trapped ions. Feedforward control, albeit being unable to correct for temporal correlations in the noise, can be used to partially mitigate errors in these applications. A similar approach may ameliorate errors other than the leading order rotation error captured by our noise model, but would require the development of noise models accurately describing such errors.

In addition to feedforward control, it may be possible to incorporate feedback control to mitigate the motional noise. For example, the motional state of an an ancilla ion can be periodically measured. Such a strong measurement, or relatedly mid-circuit cooling, would restart the ions' diffusion process, effectively reducing the time $\tau$ over which the system undergoes diffusion. Knowledge of the motional state can then be used to generate feedforward corrections until they are recalibrated by the next measurement. 

Other common quantum platforms such as superconducting qubits and Rydberg atoms also suffer from unitary gate errors. The physical mechanisms underlying these errors, however, is quite different from that of trapped ions and understanding the structure of the optimal feedforward correction in these systems may provide insight into which quantum algorithms and simulations are best suited to different platforms.\\

\section*{Acknowledgments}
\label{sec:Acknowledgments}
We thank Marko Cetina and Christopher Monroe for stimulating discussions. K.S. and E.D. acknowledge funding from ARO grant number W911NF-20-1-0163, Harvard-MIT CUA, NSF EAGER-QAC-QCH award No. 2037687, and the NSF grant No. OAC-1934714. The Flatiron Institute is a division of the Simons Foundation. D.S. acknowledges funding from the Harvard Quantum Initiative Seed Funding program and AFOSR: Grant FA9550-21-1-0236. The Flatiron Institute is a division of the Simons Foundation.



%

\end{document}